\documentclass[twocolumn]{aastex62}
\usepackage{amsmath}
\usepackage{wasysym}

\graphicspath{{figs/}}

\newenvironment{tightcenter}{%
	\setlength\topsep{0pt}
	\setlength\parskip{0pt}
	\begin{center}
}{%
 	\end{center}
}

\newcommand{\angstrom}{\textup{\AA}}
\newcommand\chandra{Chandra}

\newcommand\ciao{CIAO}
\newcommand\sherpa{Sherpa}
\newcommand\pimms{PIMMS}

\newcommand\caldb{CALDB}

\newcommand\alphaox{\alpha_{\rm ox}}

\shorttitle{}
\shortauthors{Snios et al.}

\begin{document}

\title{Discovery of Candidate X-ray Jets in High-Redshift Quasars}

\author{Bradford Snios} 
\affil{Center for Astrophysics \textbar\ Harvard  \&  Smithsonian, Cambridge, MA 02138, USA}
\author{Daniel A. Schwartz} 
\affil{Center for Astrophysics \textbar\ Harvard  \&  Smithsonian, Cambridge, MA 02138, USA}
\author{Aneta Siemiginowska} 
\affil{Center for Astrophysics \textbar\ Harvard  \&  Smithsonian, Cambridge, MA 02138, USA}
\author{Ma\l{}gosia Sobolewska} 
\affil{Center for Astrophysics \textbar\ Harvard  \&  Smithsonian, Cambridge, MA 02138, USA}
\author{Mark Birkinshaw}
\affil{H. H. Wills Physics Laboratory, University of Bristol, Bristol BS8 1TL, UK}
\author{C.~C. Cheung}
\affil{Space Science Division, Naval Research Laboratory, Washington, DC 20375, USA}
\author{Doug B. Gobeille}
\affil{Physics Department, University of Rhode
Island, Kingston, RI 02881, USA}
\author{Herman L. Marshall}
\affil{Kavli Institute for Astrophysics and Space Research, Massachusetts Institute of Technology, Cambridge, MA 02139, USA}
\author{Giulia Migliori}
\affil{Department of Physics and Astronomy, University of Bologna, Via Gobetti 93/2, I-40129 Bologna, Italy}
\affil{INAF-Institute of Radio Astronomy, Bologna, Via Gobetti 101, I-40129 Bologna, Italy}
\author{John F.~C. Wardle}
\affil{Physics Department, Brandeis University, Waltham, MA 02454, USA}
\author{Diana M. Worrall}
\affil{H. H. Wills Physics Laboratory, University of Bristol, Bristol BS8 1TL, UK}

\begin{abstract}
We present \chandra\ X-ray observations of 14 radio-loud quasars at redshifts $3 < z < 4$, selected from a well-defined sample. All quasars are detected in the 0.5--7.0\,keV energy band, and resolved X-ray features are detected in five of the objects at distances of 1--12\arcsec\ from the quasar core. The X-ray features are spatially coincident with known radio features for four of the five quasars. This indicates that these systems contain X-ray jets. X-ray fluxes and luminosities are measured, and jet-to-core X-ray flux ratios are estimated. The flux ratios are consistent with those observed for nearby jet systems, suggesting that the observed X-ray emission mechanism is independent of redshift. For quasars with undetected jets, an upper limit on the average X-ray jet intensity is estimated using a stacked image analysis. Emission spectra of the quasar cores are extracted and modeled to obtain best-fit photon indices, and an Fe\,K emission line is detected from one quasar in our sample. We compare X-ray spectral properties with optical and radio emission in the context of both our sample and other quasar surveys.  
\end{abstract}


\section{Introduction}
\label{sect:intro}

Relativistic jet outflows from active galactic nuclei can transport significant energy from the central supermassive black hole to the surrounding intercluster medium \citep[ICM;][]{Scheuer1974,Scheuer1982,Begelman1984}. Interactions between the jet and ICM will create radio lobes at kiloparsec-scale distances, making radio observations well suited for extragalactic jet studies \citep[e.g.,][]{Blandford1974,Hargrave1974,Perley1984}. Jets are responsible for feedback processes that prevent central gas from cooling in clusters of galaxies \citep{Birzan2004, Rafferty2006, McNamara2007}, consequently reducing star formation at the cluster centers \citep{Fabian1994, Fabian2012}. Thus, interactions between jets and the surrounding medium govern the overall evolution of these extragalactic systems. 

\begin{table*}
	\caption{\chandra{} Observations of Quasar Sample}
	\begin{tightcenter}
	\label{table:obs}
	\begin{tabular}{ c c c c c c c c c D }
		\hline \hline
		Object & $z$\tablenotemark{a} & R.A.\tablenotemark{b} & Decl.\tablenotemark{b} & ObsID\tablenotemark{c} & Observation & $t_{\rm exp}$\tablenotemark{d} & log($M_{\rm BH}$)\tablenotemark{e} & log($L_{\rm bol}$)\tablenotemark{e} & \multicolumn{2}{c}{log($\lambda_{\rm Edd}$)\tablenotemark{f}}\\
        & & [J2000] & [J2000] & & Date & [ks] & [log($M_{\odot})$] & [log($\rm erg\,s^{-1}$)]& & \\
		\hline
	    \decimals
		J0801$+$4725 & 3.256 & 08:01:37.682 & +47:25:28.24 & 20405 & 2018 Jan 19 & 9.40 & $8.53\pm0.47$ & 46.97 & $0.34$ \\ 
		J0805$+$6144 & 3.033 & 08:05:18.180 & +61:44:23.70 & 20399 & 2018 Jun 07 & 9.75 & -- & -- & \multicolumn{2}{c}{--} \\ 
		J0833$+$0959 & 3.713 & 08:33:22.514 & +09:59:41.14 & 20401 & 2019 Feb 04 & 9.57 & $9.58\pm0.09$ & 47.10 & $-0.58$ \\
		J0909$+$0354 & 3.288 & 09:09:15.915 & +03:54:42.98 & 20404 & 2018 Mar 03 & 9.57 & $9.55\pm0.05$ & 46.88 & $-0.77$ \\
		J0933$+$2845 & 3.431 & 09:33:37.298 & +28:45:32.24 & 20403 & 2018 Jun 10 & 9.57 & $9.56\pm0.01$ & 47.27 & $-0.39$ \\
	    J1016$+$2037 & 3.114 & 10:16:44.322 & +20:37:47.30 & 20411 & 2018 Jan 24 & 9.57 & $8.91\pm0.03$ & 47.08 & $0.07$ \\
	    J1128$+$2326 & 3.042 & 11:28:51.701 & +23:26:17.35 & 20412 & 2019 Mar 10 & 9.57 & $9.08\pm0.04$ & 47.07 & $-0.11$ \\
	    J1223$+$5038 & 3.491 & 12:23:43.169 & +50:37:53.40 & 20402 & 2018 Jul 31 & 9.57 & $9.99\pm0.02$ & 47.86 & $-0.23$ \\
	    J1405$+$0415 & 3.215 & 14:05:01.120 & +04:15:35.82 & 20408 & 2018 May 08 & 9.57 & $8.86\pm0.04$ & 46.97 & $-0.01$ \\
	    J1435$+$5435 & 3.810 & 14:35:33.779 & +54:35:59.31 & 20400 & 2018 Sep 07 & 9.57 & $8.72\pm0.06$ & 46.63 & $-0.19$ \\
	    J1610$+$1811 & 3.122 & 16:10:05.289 & +18:11:43.47 & 20410 & 2018 May 24  & 9.09 & $9.94\pm0.03$ & 47.31 & $-0.73$ \\
	    J1616$+$0459 & 3.212 & 16:16:37.557 & +04:59:32.74 & 20407 & 2018 May 08 & 9.48 & -- & -- & \multicolumn{2}{c}{--} \\
		J1655$+$3242 & 3.181 & 16:55:19.225 & +32:42:41.13 & 20409 & 2018 Nov 26 & 9.57 & $8.69\pm0.03$ & 46.80 & $0.01$ \\
	    J1655$+$1948 & 3.262 & 16:55:43.568 & +19:48:47.12 & 20406 & 2018 Jun 17 & 9.57 & $9.06\pm0.04$ & 46.78 & $-0.38$ \\
		\hline
	\end{tabular}
	\end{tightcenter}
	\tablenotemark{a}{Redshift measurements from \cite{Sowards-Emmerd2005}, \cite{Husband2015}, and \cite{Paris2018}.}
	\tablenotemark{b}{Radio centroid coordinates from VLA positions reported in \cite{Gobeille2014}.}
	\tablenotemark{c}{Observations performed using \chandra\ ACIS-S instrument with the aimpoint on the S3 chip.}
	\tablenotemark{d}{Total exposure time after flare removal reprocessing and dead time correction.}
	\tablenotemark{e}{Measurements from \cite{Rakshit2020}.}
	\tablenotemark{f}{Eddington ratios for the quasar sample.}
\end{table*}

X-ray observations of extragalactic radio sources provide insights into fundamental physical processes present within these systems \citep[e.g.,][]{Heinz1998, Harris2006, Stawarz2008, Worrall2008}. Although the origin of the X-ray jet emission mechanism is not uniquely defined, one probable mechanism is that the X-rays are generated from inverse Compton upscattering of the cosmic microwave background radiation \citep[IC/CMB;][]{Tavecchio2000, Celotti2001}. Under such an assumption, X-ray observations of jets may be used to measure the enthalpy flux, or ``power," transported by the jets to the radio lobes and the ICM \citep{Scheuer1974, Heinz1998, Reynolds2001}.  IC/CMB is predicted to be the dominant X-ray emission mechanism for high-redshift radio jets as the cosmological diminution of surface brightness by the factor $(1+z)^{-4}$ is offset by the $(1+z)^{4}$ increase in the CMB energy density. IC/CMB is additionally bolstered by the longer lifetimes of the 100\,MeV electrons, which generate such emission, relative to the 10\,GeV electrons required for radio synchrotron radiation. These emission properties make X-ray observations well suited for detecting high-redshift jets. 

Although X-rays are uniquely suitable for investigating high-redshift jets, the only telescope presently capable of resolving extended X-ray structures from high-redshift radio sources is the \chandra{} X-ray Observatory. Previous \chandra{} observations have demonstrated this capability by resolving jets in luminous radio sources at redshifts up to $z = 4.7$ \citep[e.g.,][]{Siemiginowska2003,Cheung2012, McKeough2016, Simionescu2016}. However, high-redshift X-ray jets remain vastly undersampled in comparison to nearby sources \citep{Worrall2020}. It is therefore the focus of this work to examine \chandra{} observations of additional high-redshift radio sources for evidence of X-ray structure(s) and to quantify the emission properties of these features, particularly in the context of IC/CMB radiation. 

This paper is one in a series on \chandra{} observations of 14 radio-luminous quasars selected from GHz surveys where all targets are in the redshift range $3.0 < z < 4.0$ \citep{Schwartz2020}. Here, we investigated for X-ray emission from the sources while also analyzing the spectroscopic properties of both the quasar core and the extended features. The remainder of the paper is arranged as follows. Section~\ref{sect:observation} describes the sample selection criteria and X-ray data reprocessing. Section~\ref{sect:spec} describes the X-ray spectroscopic analysis of the sample, while Section~\ref{sect:extend} covers the X-ray morphological analysis. Section~\ref{sect:outliers} details the measured properties of the five sources from our sample where extended X-ray structure is detected, including relevant flux and surface brightness results. Optical and radio properties of the quasar cores with regard to both the sample and other quasar surveys are provided in Sections~\ref{sect:ox} and \ref{sect:radio}, respectively. Our concluding remarks are provided in Section~\ref{sect:conclude}. 

For this paper, we adopted the cosmological parameters $H_{0} = 70\rm\,km\,s^{-1}\,Mpc^{-1}$, $\Omega_{\Lambda} =0.7$, and $\Omega_{M} = 0.3$ \citep{Hinshaw2013}. 
 
\begin{figure*}
  \begin{tightcenter}   
    \includegraphics[width=0.996\linewidth]{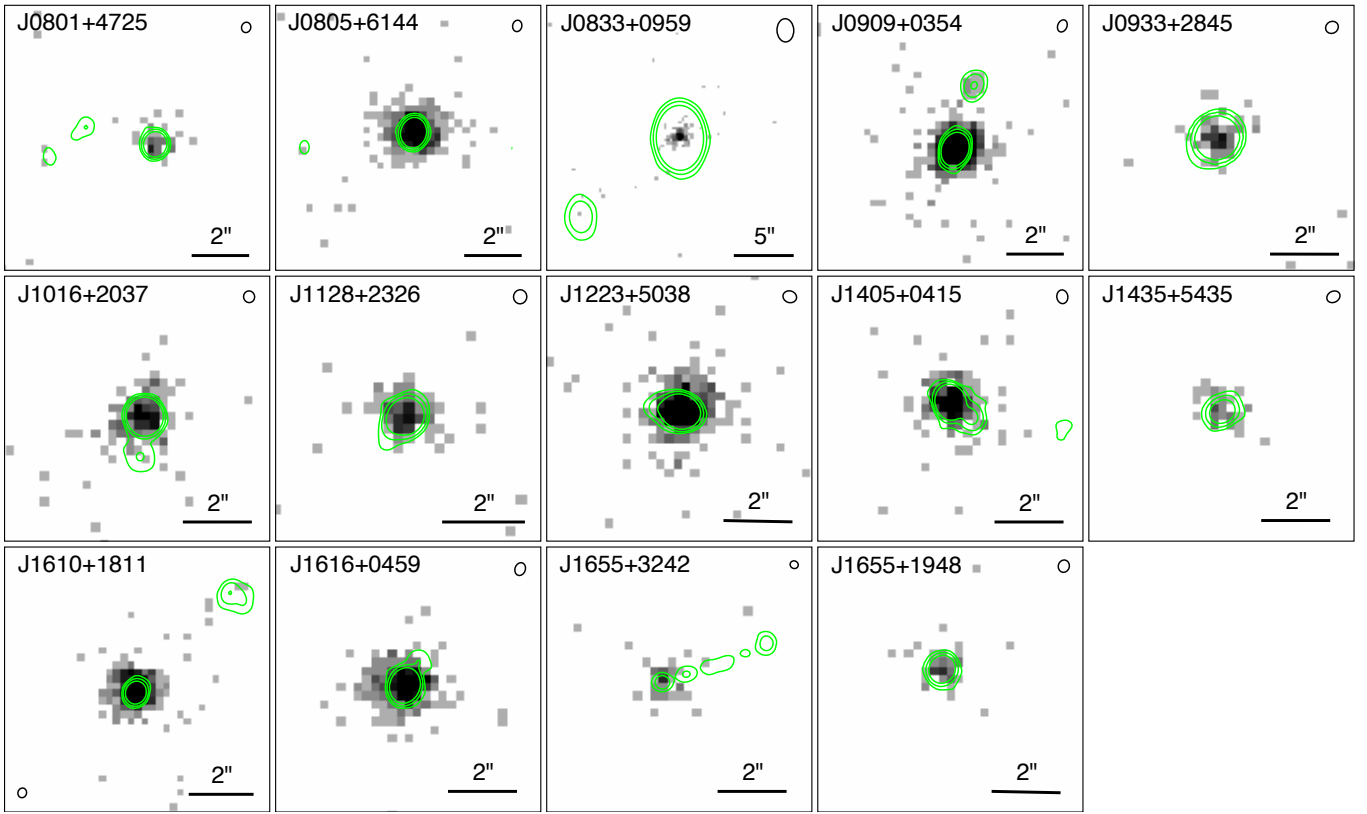}  
  \end{tightcenter}
\caption{\chandra{} 0.5--7.0\,keV images of the fourteen targets binned in 0\farcs25 pixels. The X-ray images are overlaid with radio map contours (green) from VLA observed at 6.2\,GHz, except for J0833$+$0959 which is overlaid with a 1.4\,GHz VLA radio map. The restoring beam for each radio map is shown as a black ellipse.}
\label{fig:sample}
\end{figure*}

\section{Sample Selection and Data Reduction} 
\label{sect:observation}

Targets for our analysis were selected from a catalog of 123 radio-bright quasars at redshifts $z>2.5$ \citep{Gobeille2011,Gobeille2014} constructed from an overlapping region of the VLA-FIRST radio survey \citep{Becker1995} and the Sloan Digital Sky Survey \citep{Abazajian2003}. All sources in the catalog have a spectroscopically measured redshift and a flux density in excess of 70\,mJy at either 1.4 or 5\,GHz. From this catalog, we focused on the 61 sources that possess resolved radio features at 1\arcsec{} or better resolution. We also prioritized sources where the separation distance between the radio features exceeded 1\arcsec{} as such spatial separations are resolvable with the \chandra\ X-ray Observatory, assuming comparable sizes for any X-ray counterparts. Lastly, we eliminated radio objects identified as triples because their jets are less likely to be at a line of sight that will achieve the relativistic beaming required for detection in X-rays. Of the remaining 31 sources that satisfied these criteria, 14 objects at redshifts $z > 3$ with resolved radio structure that had previously not been observed in X-rays were selected for our survey. 

Each target selected for our survey was observed with \chandra\ using the Advanced CCD Imaging Spectrometer (ACIS) with the aimpoint centered on the S3 chip. The instrument was placed in the 1/4 subarray timed exposure mode with {\tt vfaint} telemetry enabled in an effort to mitigate pileup. Roll direction for each observation was defined such that the radio features did not coincide with the readout direction of the chip. We also confirmed that the known \chandra\ PSF artifact\footnote{See the PSF artifact caveat in the \ciao{} User Guide: \\ \url{https://cxc.harvard.edu/ciao/caveats/psf_artifact.html}}, which can create nonphysical X-ray features on a subarcsecond scale, was not coincident with any radio feature in our sample. 

\begin{table*}
	\caption{VLA Observations of Quasar Sample}
	\label{table:radio_obs}
	\begin{tightcenter}
	\begin{tabular}{ccccccDc}
		\hline \hline
		Object & Exposure & Frequency & Observation & Beam & Beam & \multicolumn2c{Beam Position} & Minimum Contour \\
		& [s] & [GHz] & Date & Maximum & Minimum & \multicolumn2c{Angle} & Intensity  \\
		& & &  & [arcsec] & [arcsec] & \multicolumn2c{[deg]} & [mJy beam$^{-1}$] \\
		\hline
		\decimals
        J0801$+$4725 & 378.9 & 6.2 & 2012 Nov 18 &  0.36 & 0.31 & \hspace{1.3em}-16.0 & 0.45 \\
        J0805$+$6144 & 319.1 & 6.2 & 2012 Nov 18 & 0.40 & 0.32 & -11.5 & 0.68 \\
        J0833$+$0959 & 580.0 & 1.4 & 2008 Oct 23 & 1.95 & 1.41 & 2.3 & 0.37 \\
        J0909$+$0354 & 319.1 & 6.2 & 2012 Nov 18 & 0.44 & 0.32 & -23.5 & 0.10 \\
        J0933$+$2845 & 319.1 & 6.2 & 2012 Nov 18 & 0.37 & 0.34 & -57.9 & 0.52 \\
        J1016$+$2037 & 398.9 & 6.2 & 2012 Nov 12 & 0.35 & 0.32 & 11.4 & 0.13 \\
        J1128$+$2326 & 319.1 & 6.2 & 2012 Nov 12 & 0.34 & 0.32 & -7.7 & 0.15 \\
		J1223$+$5038 & 299.2 & 6.2 & 2012 Nov 12 & 0.41 & 0.33 & 72.3 & 0.12 \\
        J1405$+$0415 & 319.1 & 6.2 & 2012 Nov 18 & 0.39 & 0.32 & 7.5 & 0.62 \\
        J1435$+$5435 & 319.1 & 6.2 & 2012 Nov 18 & 0.41 & 0.32 & -61.4 & 0.40 \\
        J1610$+$1811 & 319.1 & 6.2 & 2012 Nov 08 & 0.34 & 0.31 & -16.6 & 0.18 \\
        J1616$+$0459 & 259.2 & 6.2 & 2012 Nov 08 & 0.39 & 0.31 & -19.7 & 0.85 \\
        J1655$+$3242 & 378.9 & 6.2 & 2012 Nov 08 & 0.24 & 0.21 & 63.8 & 0.68 \\
        J1655$+$1948 & 319.1 & 6.2 & 2012 Nov 08 & 0.35 & 0.31 & -28.9 & 0.34 \\
		\hline
	\end{tabular}
	\end{tightcenter}
\end{table*}

All X-ray observations were analyzed using the level 2 data products from the standard \chandra\ data processing pipeline together with the software analysis package \ciao{}v4.12 with \caldb{}v4.9.2.1. Each observation was reprocessed using the routine {\tt deflare} to remove background flaring periods from the data, and the average cleaned exposure time per target is 9.53\,ks. Pileup was estimated for each source using \pimms{}v4.11a, and all sources had $< 3\%$ pileup over the 0.5--7.0\,keV energy band. Thus, pileup was ignored for our X-ray analysis. Details on our \chandra\ observations are provided in Table~\ref{table:obs}.

In addition to the X-ray data, radio observations of the sample were included in our analysis. We obtained new Karl G. Jansky Very Large Array (VLA) A-array observations of these 14 quasars (see Table~\ref{table:radio_obs} for a summary) as part of a more extensive radio imaging and follow-up program of this sample \citep{Gobeille2014}. The data were calibrated and imaged using standard procedures in CASA and AIPS. The majority of the VLA maps (13/14) were obtained as single $\sim$\,4--6 minute snapshot observations in November 2012 (program 12B-230) using two 1\,GHz wide intermediate-frequency bands centered at 4.9 and 7.4\,GHz (effective center frequency of 6.2 GHz). The maps of J1405$+$0415 and J1610$+$1811 were previously published in \cite{Schwartz2020}. For J0833$+$0959, we used an earlier VLA A-array 1.4\,GHz map we obtained in October 2008 (program AW748) and published in \cite{Gobeille2014}. 

The X-ray and radio observations were aligned based on the centroid position of the core. Centroid position was measured in each observation using the {\tt dmstat} routine in \ciao, and the  0.5--7.0\,keV X-ray observation coordinates were shifted with {\tt wcs\_update} to agree with the radio data. The average astrometric translations for the sample were $\Delta x_{\rm avg} = 0\farcs62$ and $\Delta y_{\rm avg} = 0\farcs45$, which are consistent with the overall 90\% absolute position uncertainty of 0\farcs8 for \chandra. We also generated radio contours from the available maps. The minimum contour level was defined as $\sim$\,3--5 times the rms noise measured in an off-source region on the image, and the contour levels were drawn at increasing factors of four intervals for all sources. The resulting X-ray--radio overlays are shown in Figure~\ref{fig:sample}, where the 0.5--7.0\,keV \chandra\ images are subpixel binned in 0\farcs25 pixels. Additional details on the radio maps and contours used in our analysis are provided in Table~\ref{table:radio_obs}. 
 
\section{X-ray Emission Spectra} 
\label{sect:spec}

\begin{table*}
	\caption{X-ray Properties of the Quasar Sample}	
    \label{table:xray}
    \begin{tightcenter}
	\begin{tabular}{ c c c c D c c c D D D }
		\hline \hline
		Object & $C_{\rm obs}$ & Model & $N_{\rm H}$ & \multicolumn2c{$N_{\rm H}^i$} & $\Gamma$ & $E_{\rm line}$ & $I_{\rm line}$ & \multicolumn2c{$f_{\rm 0.5-7.0\,keV}$} & \multicolumn2c{$L_{\rm 2-10\,keV}$} & \multicolumn2c{$\ell_{\rm 2\,keV}$} \\
		(1) & (2) & (3) & (4) & \multicolumn2c{(5)} & (6) & (7) & (8) & \multicolumn2c{(9)} & \multicolumn2c{(10)} & \multicolumn2c{(11)} \\
		\hline
		\decimals
        J0801$+$4725 & 48 & A & 0.0454 & $<15.5$ & $2.01\substack{+0.30\\-0.29}$ & -- & -- & $5.8\substack{+1.0\\-0.9}$ & $3.4\substack{+1.2\\-1.0}$ & $4.3\substack{+3.1\\-1.9}$ \\
		J0805$+$6144 & 475 & A & 0.0452 & $<4.7$ & $1.24\substack{+0.09\\-0.09}$ & -- & -- & $66.1\substack{+3.4\\-3.4}$ & $19.4\substack{+2.4\\-2.2}$ & $12.7\substack{+2.8\\-2.3}$ \\
		J0833$+$0959 & 173 & A & 0.0394 & $<11.5$ & $1.60\substack{+0.16\\-0.16}$ & -- & -- & $23.3\substack{+2.0\\-1.9}$ & $13.7\substack{+3.1\\-2.7}$ & $12.5\substack{+5.1\\-3.8}$ \\
		J0909$+$0354 & 797 & A & 0.0347 & $<3.5$ & $1.16\substack{+0.07\\-0.07}$ & -- & -- & $116.4\substack{+4.4\\-4.6}$ & $37.0\substack{+3.6\\-3.4}$ & $22.4\substack{+3.8\\-3.4}$ \\
		J0933$+$2845 & 80 & A & 0.0191 & $<10.7$ & $1.65\substack{+0.22\\-0.22}$ & -- & -- & $10.4\substack{+1.2\\-1.2}$ & $5.3\substack{+1.6\\-1.3}$ & $5.1\substack{+2.9\\-1.9}$ \\
	    J1016$+$2037 & 196 & A & 0.0242 & $<6.2$ & $1.67\substack{+0.14\\-0.14}$ & -- & -- & $25.6\substack{+2.1\\-2.0}$ & $10.9\substack{+2.0\\-1.8}$ & $10.6\substack{+3.6\\-2.8}$ \\
	    J1128$+$2326 & 91 & A & 0.0134 & $<9.3$ & $1.48\substack{+0.20\\-0.20}$ & -- & -- & $12.7\substack{+1.6\\-1.3}$ & $4.5\substack{+1.3\\-1.0}$ & $3.7\substack{+2.0\\-1.3}$ \\
	    J1223$+$5038 & 464 & A & 0.0169 & $<4.3$ & $1.62\substack{+0.08\\-0.09}$ & -- & -- & $62.4\substack{+3.3\\-3.1}$ & $32.7\substack{+4.1\\-3.7}$ & $30.6\substack{+6.6\\-5.5}$ \\
	    ... & ... & B & ... & $<4.3$ & $1.50\substack{+0.12\\-0.06}$ & $6.28\substack{+0.05\\-0.09}$ & $5.44\substack{+1.99\\-1.83}$ & $60.5\substack{+3.2\\-3.1}$ & $28.8\substack{+4.5\\-2.8}$ & $24.1\substack{+7.0\\-3.5}$ \\
	    J1405$+$0415 & 280 & A & 0.0217 & $<4.9$ & $1.49\substack{+0.11\\-0.11}$ & -- & -- & $37.3\substack{+2.7\\-2.3}$ & $14.8\substack{+2.4\\-2.1}$ & $12.3\substack{+3.5\\-2.7}$ \\
	    J1435$+$5435 & 38 & A & 0.0127 & $<34.0$ & $1.32\substack{+0.32\\-0.32}$ & -- & -- & $5.3\substack{+1.1\\-0.9}$ & $2.5\substack{+1.5\\-1.0}$ & $1.8\substack{+2.0\\-1.0}$ \\
	    J1610$+$1811 & 397 & A & 0.0362 & $<4.5$ & $1.64\substack{+0.10\\-0.10}$ & -- & -- & $53.4\substack{+3.0\\-2.8}$ & $22.2\substack{+2.8\\-2.5}$ & $21.1\substack{+4.8\\-4.0}$ \\
	    J1616$+$0459 & 386 & A & 0.0475 & $<6.0$ & $1.48\substack{+0.10\\-0.10}$ & -- & -- & $51.1\substack{+3.3\\-3.0}$ & $20.1\substack{+2.9\\-2.5}$ & $16.5\substack{+4.1\\-3.3}$ \\
	    J1655$+$3242 & 41 & A & 0.0222 & $<19.3$ & $1.51\substack{+0.31\\-0.31}$ & -- & -- & $5.5\substack{+1.0\\-0.9}$ & $2.2\substack{+1.0\\-0.7}$ & $1.8\substack{+1.7\\-0.9}$ \\
	    J1655$+$1948 & 47 & A & 0.0549 & $<14.1$ & $1.91\substack{+0.29\\-0.29}$ & -- & -- & $5.8\substack{+1.0\\-0.8}$ & $3.2\substack{+1.2\\-0.9}$ & $3.8\substack{+2.7\\-1.7}$ \\
		\hline
	\end{tabular}
	\end{tightcenter}
	 (1) Object name. (2) Observed counts over the 0.5--7.0\,keV band from 1\farcs5 radius circle centered on the quasar, where the average background contribution is 0.17\% of the observed counts. (3) Spectral model. (4) Galactic column density extrapolated from \cite{Dickey1990}, in units of $10^{22}\rm\,cm^{-2}$. (5) Intrinsic column density 3$\sigma$ upper limits, in units of $10^{22}\rm\,cm^{-2}$. (6) Photon index estimated from 0.5--7.0\,keV best-fit spectral model. (7) Rest-frame iron emission line energy, in unit of keV. (8) Observed line intensity, in units of $10^{-6}\rm\,photons\,cm^{-2}\,s^{-1}$. (9) Observed 0.5--7.0\,keV flux, in units of $10^{-14}\rm\,erg\,cm^{-2}\,s^{-1}$. (10) Rest-frame 2--10\,keV luminosity,  in units of $10^{45}\rm\,erg\,s^{-1}$. (11) Rest-frame, monochromatic luminosity at 2\,keV, in units of $10^{27}\rm\,erg\,s^{-1}\,Hz^{-1}$.
\end{table*}

The unabsorbed X-ray fluxes and luminosities for the sample were determined by modeling the spectrum from each source. An emission spectrum from each observation was extracted using the {\tt specextract} routine in \ciao. We defined the source region as a 1\farcs5 radius circle centered on the quasar, while the background region was defined as an annulus, also centered on the quasar, with an inner radius of 15\arcsec{}  and an outer radius of 30\arcsec. We masked any point sources in the background region, if present, to avoid count biases. We note that the average percent difference between the total and background-corrected counts is 0.17\% for our extracted spectra, so the background contribution is minor. Each spectrum was binned at 1 count per bin over the 0.5--7.0\,keV band.  

While performing our spectral analysis of the sample, we additionally tested for the presence of spectral lines, such as an Fe\,K line. We defined two different spectral models, one with an emission line and one without, and assessed both models with a likelihood ratio test for each observation. The line emission component was included in our model in cases where its introduction improved the $p$-value of the likelihood ratio test to a value less than 0.001. The models are as follows: 

Model A: {\tt phabs$\cdot$powerlaw}. This is the default model for X-ray emission from the quasar core that was used for all 14 sources. 

Model B: {\tt phabs$\cdot$(powerlaw\,$+$\,zgauss)}. This model includes the primary X-ray emission and a Gaussian emission line, where we fixed the line width to 0.001\,keV. This model was tried for all sources but preferred only for J1223$+$5038.

Each extracted spectrum was fit over the 0.5--7.0\,keV energy band using WStat in \ciao's modeling and fitting package \sherpa. Galactic hydrogen column density $N_{\rm H}$ was fixed to extrapolated values from \cite{Dickey1990}, while the photon index $\Gamma$ ($dN/dE \propto E^{-\Gamma}$) and the normalization(s) were allowed to vary for all models. Once a best-fit model was realized, an additional intrinsic absorption component (i.e., {\tt zphabs}) was added to the model to determine if the intrinsic absorption column density $N_{\rm H}^i$ could be constrained. The revised model was fit to the data with $N_{\rm H}^i$, normalization, and $\Gamma$ as free parameters. In all cases, only an upper limit on the intrinsic absorption could be established as its addition did not improve the overall fit statistics. Model parameter best-fit results from the spectral analysis and their respective $1\sigma$ confidence intervals are provided in Table~\ref{table:xray}. We note that the upper limits on $N_{\rm H}^i$ are reported at $3\sigma$ confidence. 

\begin{figure}
    \begin{tightcenter}
    \includegraphics[width=0.99\linewidth]{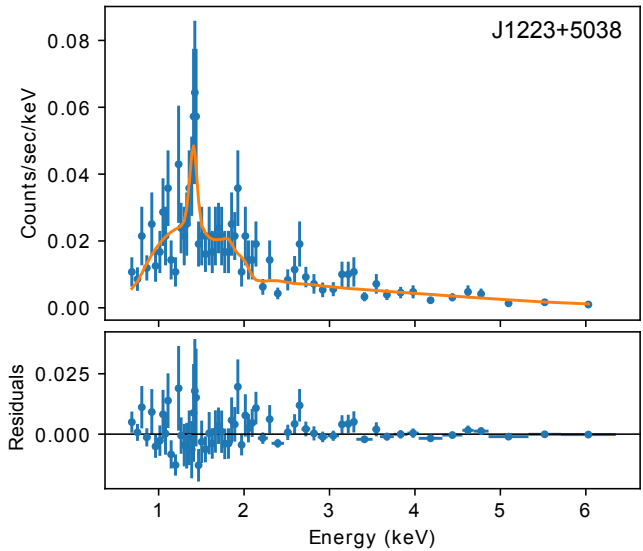} 
    \end{tightcenter}
\caption{\chandra\ X-ray spectrum of J1223$+$5038. The spectrum is binned by 5 counts per bin (solely for illustrative purposes) and is fitted over the 0.5--7.0\,keV energy range with an absorbed power-law model ($\Gamma = 1.50\substack{+0.12\\-0.06}$) and an Fe\,K emission line at the rest energy of 6.28\,keV, or observed energy of 1.39\,keV. The lower panel shows the residuals from the best-fit model.}
\label{fig:spectrum}
\end{figure}

We found from our spectral analysis that all 14 sources in the sample could be individually fit with Model A, providing constraints on both $N_{\rm H}^i$ and $\Gamma$. Additionally, we determined from our likelihood ratio tests that J1223$+$5038 was best fit with Model B due to the presence of an emission line in its spectrum. The measured  emission line is consistent with neutral Fe emission at 6.4\,keV, while the $N_{\rm H}^i$ and $\Gamma$ best-fit parameters agree with the Model A results. The observed spectrum and model fit for J1223$+$5038 are shown in Figure~\ref{fig:spectrum}, and the best-fit results are in Table~\ref{table:xray}. 

Once the best-fit spectral models were obtained, we measured the observed Galactic absorption-corrected 0.5--7.0\,keV flux $f_{\rm 0.5-7.0\,keV}$ for each source. The rest-frame 2--10\,keV luminosity $L_{\rm 2-10\,keV}$ and monochromatic 2\,keV luminosity $\ell_{\rm 2\,keV}$ were additionally determined from each best-fit model. We applied aperture correction for all measured fluxes and luminosities, where the correction factor was derived from our encircled counts fraction (ECF) analysis of the sources (see Section~\ref{sect:extend}). The average correction factor for our sample is 1.075, and the range is 1.056--1.098. All measured X-ray fluxes and luminosities for the sample are shown in Table~\ref{table:xray}. 
\section{X-ray Morphology}
 \label{sect:extend}

As discussed in Section~\ref{sect:observation}, our sample of high-redshift quasars has known radio features at distances of 1\arcsec{} or greater from the quasar core. Existing X-ray counterparts to these radio features may be resolvable with our \chandra\ observations, assuming that both a satisfactory signal-to-noise ratio (S/N) is achieved and the radio-/X-ray-emitting regions are comparable in size. Thus, we investigated our sample for evidence of resolved X-ray structures. 

\begin{figure}
    \begin{tightcenter}
    \includegraphics[width=0.99\linewidth]{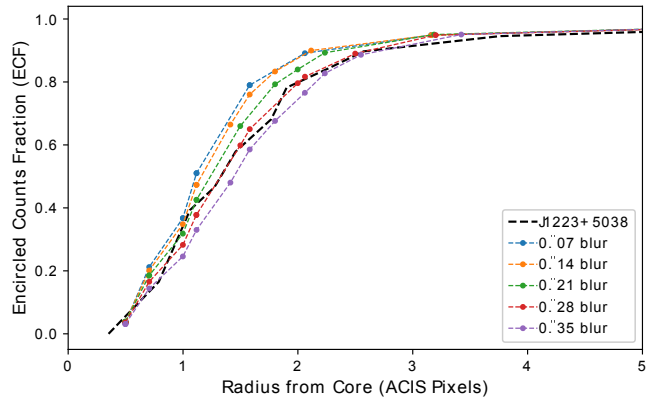} 
    \end{tightcenter}
\caption{A comparison of the background-subtracted encircled counts fraction (ECF) for the quasar J1223$+$5038 versus simulated PSFs generated from the {\tt simulate\_psf} routine in \ciao. For each simulation, a different aspect blur parameter was used. We found that a blur parameter of 0\farcs28 is best for reproducing the sources in our sample.}
\label{fig:ecf}
\end{figure}

The asymmetric point-spread function (PSF) for \chandra\footnote{See `Understanding the Chandra PSF' thread in the \ciao{} guide:  \url{https://cxc.cfa.harvard.edu/ciao/PSFs/psf_central.html}} 
must be considered when investigating features on scales of $\sim$\,1\arcsec, which we suspected for our sources. As there is presently no analytic PSF for \chandra, we generated synthetic PSF images for our observations. We began by simulating 500 ray-tracing files for each X-ray observation using ChaRT\,v2, a web interface to the SAOsac ray-trace code. The ray-tracing files and various physical parameters of the observation, such as exposure time and detector orientation, were input into the {\tt simulate\_psf} script in \ciao\ to accurately generate simulated PSF images. All ray-tracing files were projected using MARX\,v5.5.0, and the extracted spectra from Section~\ref{sect:spec} were used to reproduce the spectral response of our observations. Since the default value for the aspect blur parameter in {\tt simulate\_psf} has previously been shown to generate a simulated PSF profile narrower than what is observed,\footnote{See `The AspectBlur Parameter in MARX' thread in the \ciao{} guide:  \url{https://cxc.cfa.harvard.edu/ciao/why/aspectblur.html}}, we generated multiple simulated PSF with different aspect blur values. The ECF was calculated for each simulation and compared with the observed data. See Figure~\ref{fig:ecf} for an example of this comparison for J1223$+$5038. From our analysis, we found an aspect blur value of 0\farcs28 to best fit all sources in our survey. 

\begin{figure}
    \begin{tightcenter}
    \includegraphics[width=0.999\linewidth]{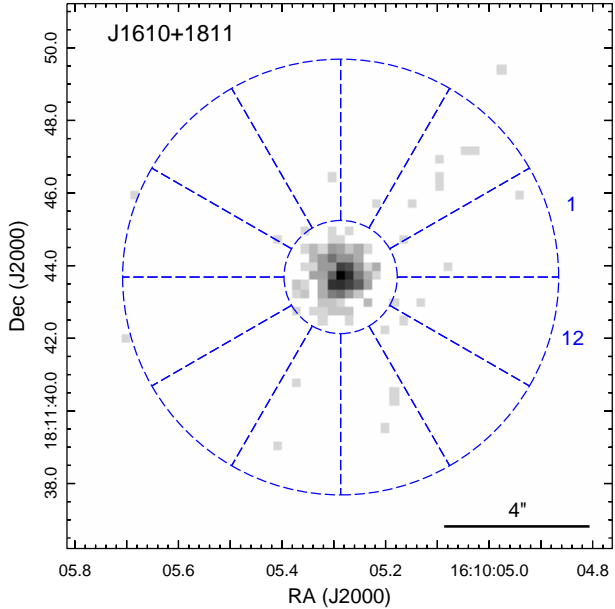} 
    \end{tightcenter}
\caption{The annular sectors defined in Section~\ref{sect:extend} for 1 of the 14 sources in our quasar sample. Each annulus was divided into twelve 30$^{\circ}$ sectors, beginning West and rotating counterclockwise. In the source shown, the inner and outer radii correspond to 1\farcs5 and 6\arcsec, respectively. For convenience, sectors 1 and 12 have been labeled.}
\label{fig:sectors}
\end{figure}

After verifying that all PSF simulations were caligned with their respective \chandra\ observations to subpixel accuracy, we defined an annulus region for each quasar. The outer annular radius was set such that it encompassed the observed radio features plus an additional 1\arcsec{} to account for \chandra\ PSF blurring. Because the radio map for J0833$+$0959 has a poorer resolution than the remainder of our sample, we defined its outer annular radius as 1\arcsec{} from the centroid of the external radio feature. The inner annular radius was set equal to the 95\% ECF radius of the core, as measured from our simulated ECF profiles, in cases where the core separation distance for the nearest radio feature is $>2\arcsec{}$. In cases where the core separation distance is $\leq 2\arcsec$, the inner radius was set equal to 85\% ECF radius. The resulting inner radii for the sample ranged between 1\farcs0--1\farcs7, and the outer radii were between 2\farcs0--11\farcs0. Each annulus was additionally divided into twelve $30{^\circ}$ sectors, where we defined our coordinate system as $0{^\circ}$ West rotating counterclockwise. The sectors used for J1610$+$1811 are shown in Figure~\ref{fig:sectors}. 

Total counts for each annular sector of the observations and simulations were measured using {\tt dmextract} in \ciao. Counts were also measured for a circular region 1\arcsec{} in radius surrounding the core in each X-ray image, and the simulated leakage counts into each annular sector were scaled based on the ratio of the observed-to-simulated core counts.  The average observed background counts per sector were estimated using the background regions discussed in Section~\ref{sect:spec}. 

\subsection{Statistical Assessment of X-ray Features}
\label{sect:annulus}

Having measured the counts per annular sector for both observations and simulations, we examined which sectors, if any, have elevated observed counts relative to the simulated PSF as this would indicate the presence of extended X-ray emission. To ensure a rigorous detection criterion, we derived a counts probability for each sector based on the simulated and field background emissions in order to find sectors that are statistically significant outliers. 

To begin, we inferred that the simulated count probability distribution of each sector in our analysis could be accurately modeled with Poisson statistics. Thus, the simulated counts per sector may be represented as a Poisson distribution where the average is calculated from the 500 simulations. We note that the simulations do not include background emission, which must be accounted for when comparing to the observed counts. The background counts may similarly be represented as another Poisson distribution where the mean value corresponds to the predicted background counts per sector. Because the sum of independent Poisson random variables is itself Poisson \citep{Grimmett1986}, we may define the sum of simulated and background distributions as $P(\lambda_{1}+\lambda_{2}$), where $\lambda_{1}$ is the mean simulated counts per sector and $\lambda_{2}$ is the expected background counts per sector. We may therefore accurately determine the probability for the total observed counts per annular sector for each source. 

\begin{figure*}
  \begin{tightcenter}   
    \includegraphics[width=0.98\linewidth]{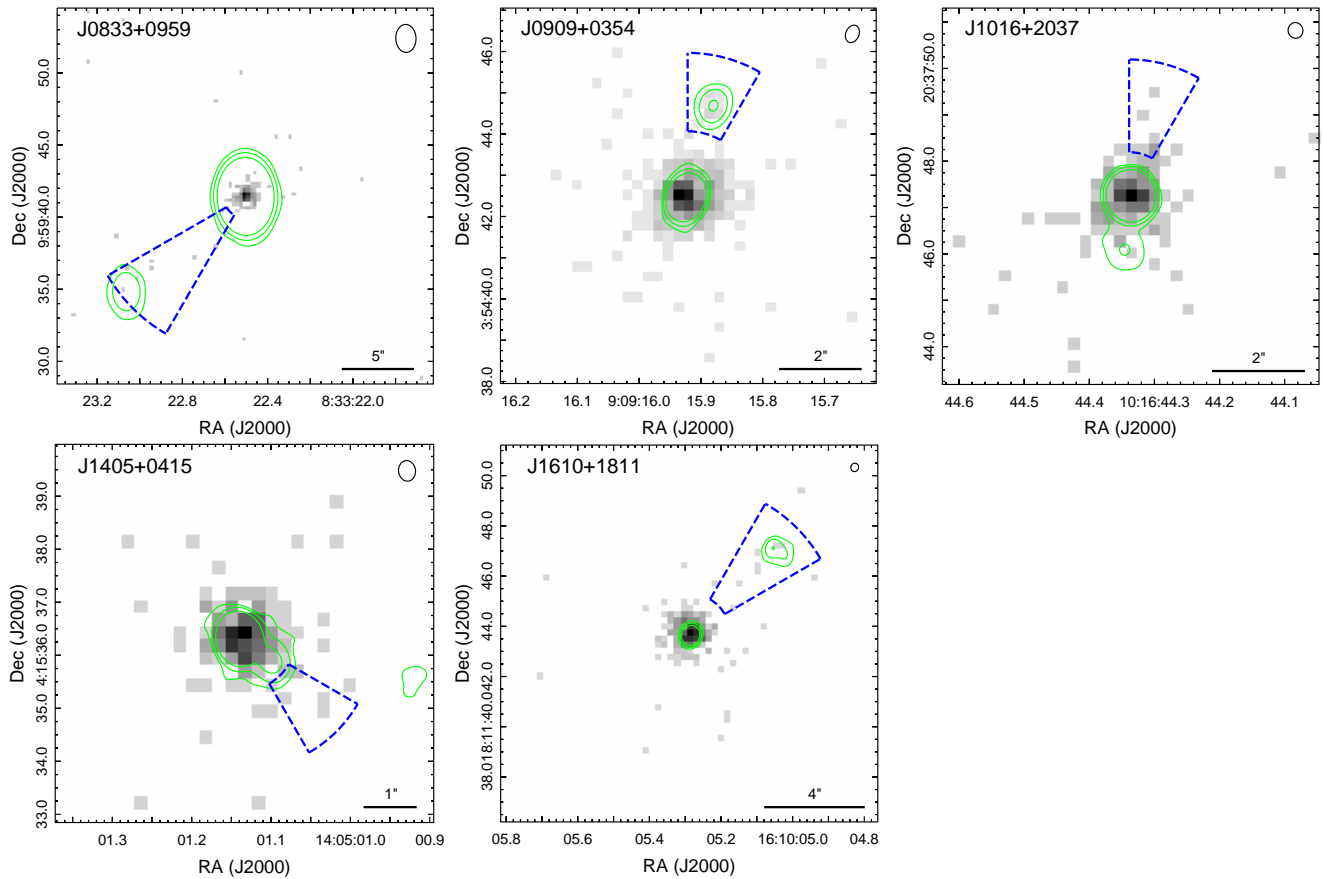}
  \end{tightcenter}
\caption{\chandra{} 0.5--7.0\,keV images of the five sources where resolved X-ray structure is detected from our statistical analysis. Each image is binned in 0\farcs25 pixels and overlaid with radio map contours (green). The regions shown are the sectors where X-ray features were detected, and the restoring beam for each radio map is shown as a black ellipse.}
\label{fig:extended}
\end{figure*}

Using the predicted probability distributions, we calculated the cumulative probability of obtaining the observed count rate or higher for each sector of each source. Due to our method of defining sectors, it may be possible that extended X-ray features lie in multiple sectors. We therefore also assessed probabilities for the sum of adjacent sectors, up to a total sector size of $90^{\circ}$, in order to address these edge cases. We defined a probability $p \leq 0.005$ as the detection threshold for the sample. Based on our defined significance threshold, we found five sectors from five unique sources that have significant evidence of X-ray emission. The X-ray features and their respective probabilities are provided in Table~\ref{table:extend}, and images of the five sources with their corresponding emission sectors are shown in Figure~\ref{fig:extended}. A machine-readable table of the complete sector analysis for all sources is available as an accompanying online resource to this article.

\section{Resolved X-ray Features} 
\label{sect:outliers}

The morphological analysis in Section~\ref{sect:extend} identified five quasars with extended X-ray emission at distances of 1--12\arcsec\ from the quasar core, where the projected jet length range is 20--80\,kpc. These jet lengths are consistent with those observed for nearby, high-luminosity jets \citep{Marshall2018}. Examination of the X-ray properties of the quasar cores, as well as the multiwavelength properties of the cores (see Sections~\ref{sect:ox} and \ref{sect:radio}), show no identifiable characteristics that would isolate these sources from the remainder of the sample. We therefore sought to quantify the X-ray properties of the five extended X-ray features.

\begin{table}
	\caption{X-ray Morphological Analysis
	\label{table:extend}}
	\begin{tightcenter}
	\footnotesize
	\begin{tabular}{c c D c c c D}
		\hline \hline
		Object & $r_{\rm in}$ & \multicolumn2c{$r_{\rm out}$} & $C_{\rm obs}$ & $C_{\rm PSF}$ & $C_{\rm bg}$ & \multicolumn2c{$p$} \\
		(1) & (2) & \multicolumn2c{(3)} & (4) & (5) & (6) & \multicolumn2c{(7)} \\
		\hline
		J0833$+$0959 & 1.5 & 11.0 & 8 & 0.84 & 0.60 & $<0.0001$ \\ 
		J0909$+$0354 & 1.7 & 3.5 & 12 & 2.42 & 0.07 & $<0.0001$ \\ 
		J1016$+$2037 & 1.0 & 3.0 & 8 & 1.67 & 0.04 & $0.0004$ \\ 
		J1405$+$0415 & 1.0 & 2.5 & 8 & 2.19 & 0.03 & $0.0021$ \\ 
		J1610$+$1811 & 1.6 & 6.0 & 7 & 1.68 & 0.21 & $0.0033$ \\ 
		\hline
	\end{tabular}
    \end{tightcenter}
    {(1) Object name. (2) Inner annular radius, in units of arcseconds. (3) Outer annular radius, in units of arcseconds. (4) Observed 0.5--7.0\,keV counts. (5) Mean 0.5--7.0\,keV counts from simulated PSF. (6) Expected background 0.5--7.0\,keV counts. (7) Cumulative Poisson probability of detecting counts $\geq C_{\rm obs}$. The results listed in the table are those sectors where the probability was $\leq$ 0.005, indicating strong evidence of a resolved X-ray feature. (A machine-readable table of the complete sector analysis for all sources is available.)}
\end{table}

\subsection{Flux and Surface Brightness}
\label{sect:flux_extend}

The flux and the surface brightness of each X-ray feature detected in Section~\ref{sect:annulus} may be measured, given some assumptions on the X-ray emission. Due to the limited X-ray counts in our observations, identification of edges for the X-ray extension could not be performed using either surface brightness profiles or contour mapping, as is standard in jet and hotspot analyses. We instead assumed that the X-rays were generated from jets that extended from the quasar core. We additionally assumed that the annular sectors in Figure~\ref{fig:sectors} were sufficient in size to encompass the extended X-ray emission. The annular sectors were verified to reproduce the same measured count rates as other region shapes, such as rectangular and elliptical regions \citep[e.g.,][]{Schwartz2020}. Furthermore, we found the annular sectors to minimize the overall systematic error introduced from the PSF counts, with a 20\% reduction in $C_{\rm PSF}$ relative to similarly sized rectangles and ellipses, while also improving the overall reproducibility of our flux measurements. Thus, we used the count rates measured with the annular sectors (Table~\ref{table:extend}) for our flux analysis.

The source counts of each region were calculated by subtracting the mean counts from the simulated PSF and the expected background counts from the total observed counts. We then modeled each region with a {\tt phabs$\cdot$powerlaw} expression in \pimms{}v4.11a, using the latest \chandra\ response files, and set the spectral normalization equal to our X-ray jet count rates. The Galactic column density was fixed equal to values from \cite{Dickey1990}. We also fixed the photon index to $\Gamma = 1.9$, which we inferred to be a standard slope for X-ray jet and knot spectra \citep[e.g.,][]{Marshall2002,Siemiginowska2002,Schwartz2006,Schwartz2006b,Goodger2010,Zhang2018}. The observed 0.5--7.0\,keV X-ray jet fluxes are shown in Table~\ref{table:flux}. We note that the flux values for J1405$+$0415 and J1610$+$1811 agree with previous measurements from \cite{Schwartz2020}. Rest-frame 2--10\,keV luminosities were also measured for each source and are provided in Table~\ref{table:flux}. We found the fluxes and luminosities to be similar for the five jets.  

We additionally calculated the surface brightness values for each source. We assumed the emitting area originated at the quasar core and extends to the outer radius of our defined regions. The width of the emitting region was set equal to the FWHM of \chandra, which is 0\farcs5. See Table~\ref{table:flux} for the flux, luminosity, and surface brightness measurements of the X-ray jets. Given the limited count statistics for these extended X-ray features, the measurements in Table~\ref{table:flux} have an approximate factor of 2 error.  

\subsection{IC/CMB Emission}
\label{sect:iccmb}

The IC/CMB in a jet with bulk relativistic motion is a known X-ray emission mechanism in quasar jets, where the broadband emission is attributed to a single spectrum of relativistic electrons \citep{Tavecchio2000,Celotti2001,Schwartz2002,Siemiginowska2002,Schwartz2006,Schwartz2006b,Worrall2009,Marshall2011,Marshall2018, Worrall2020}. However, recent studies have demonstrated that IC/CMB from a single electron population does not reproduce the observed spectral properties of several lower-redshift jets  \citep[e.g.,][]{Jester2006,Siemiginowska2007,Cara2010,Meyer2015,Clautice2016}. In addition, upper limits on 0.1--100\,GeV $\gamma$-ray production in jets measured from Fermi observations imply that a single electron population is insufficient to produce the observed X-ray emission in several jets at $z < 1$ \citep{Breiding2017,Meyer2017b}, though this issue may be alleviated with increased electron cooling \citep{Lucchini2016}. Overall, these results refute the single electron population required for standard IC/CMB jet models and instead require multiple electron populations and/or different emission mechanisms to reproduce the observed multiwavelength radiation.

Despite these results for low-redshift ($z<3$) systems, IC/CMB is theorized to be the predominant emission mechanism at $z \gtrsim 3$ as the energy density of the CMB exceeds the magnetic energy density for the quasar jets \citep{Schwartz2020}. This effect increases the X-ray intensity of the high-redshift jet as well as the X-ray-to-radio intensity ratio, which has been observed in a number of systems \citep{Siemiginowska2003,Cheung2004,Cheung2006,Cheung2012,McKeough2016,Simionescu2016}. Although our sample lacks the required count statistics necessary to directly confirm or deny the presence of IC/CMB via spectroscopic modeling (see Section~\ref{sect:flux_extend}), a predicted effect of the elevated IC/CMB jet emission is an increase in the jet flux $f_{\rm jet}$ relative to the quasar core flux $f_{\rm core}$. Thus, we examined the jet-to-core flux ratios from our sample for evidence of redshift dependence in the context of IC/CMB jet emission.  
\bibliographystyle{aasjournal}

We estimated the jet-to-core flux ratios $f_{\rm jet}/f_{\rm core}$ for our sample using the observed 0.5--7.0\,keV fluxes in Tables~\ref{table:xray} and \ref{table:flux}, and the ratio results are shown in Table~\ref{table:flux}. The measured flux ratios are between 1.0--3.6\%, with a mean value of 2.2\%. These results are in excellent agreement with the median 2\% $f_{\rm jet}$/$f_{\rm core}$ found by \cite{Marshall2018} for quasars at $z < 2$. The X-ray flux densities we derived consequently do not follow the expected $(1 + z)^4$ dependence for IC/CMB, which is consistent with the conclusion of \cite{Marshall2011}. Our sample also agrees with the $f_{\rm jet}/f_{\rm core}$ results from \cite{Worrall2020} that were based on the few known X-ray jets at $z>3.5$ with good S/N.

Although we did not observe an elevation in X-ray jet flux as a function of redshift, we stress that this result is indicative. It is possible that the quasar core is also dominated by beaming \citep{Worrall1987}, biasing our X-ray ratios. Additionally, the shallow line-of-sight expected for our quasars will cause the innermost region of the jet to appear as part of the core due to the limited spatial resolution of \chandra. This innermost region of the jet has been shown to produce the brightest features in some nearby sources \citep[e.g.,][]{Snios2019b,Snios2019}, which may bias our findings. Our comparison also requires that the high- and low-redshift quasar cores be physically consistent with one another, which may not be valid. Deeper \chandra\ observations are necessary to investigate the spectrum and spatial structure of these systems. For now, we can only reiterate that the flux density jet-to-core ratio derived from our high-redshift sample is consistent with measurements from low-redshift quasars, suggesting that it is independent of redshift.

\subsection{Flux Limits for Undetected X-ray Jets}
\label{sect:undected}

Despite X-ray features being detected from five sources in our quasar sample, the remaining nine sources show no evidence of resolved X-ray structure(s). However, X-ray jets should be cospatial with the radio features observed at GHz frequencies, so it is possible that X-ray emission is present but resides below our detection threshold. We therefore stacked the nine sources with no detections in an effort to measure the average intensity of their X-ray jets. 

\begin{table}
	\caption{X-ray Jet Properties
	\label{table:flux}}
	\begin{tightcenter}
	\begin{tabular}{c D c D c c c }
		\hline \hline
		Object & \multicolumn2c{$d_{\rm jet}$} & $C_{\rm src}$ & \multicolumn2c{$f_{\rm jet}$} & $S_{\rm jet}$ & $L_{\rm jet}$ & $f_{\rm jet}$/$f_{\rm core}$\\
		(1) & \multicolumn2c{(2)} & (3) & \multicolumn2c{(4)} & (5) & (6) & (7)\\
		\hline
		J0833$+$0959 & 11.0 & 6.56 & 8.3 & 1.5 & 6.2 & 0.036 \\
		J0909$+$0354 & 3.5 & 9.51 & 12.1 & 6.9	& 6.7 & 0.010 \\ 
		J1016$+$2037 & 3.0 & 6.29 & 8.0	& 5.4	& 3.9 & 0.031 \\ 
		J1405$+$0415 & 2.5 & 5.78 & 7.4	& 5.9	& 3.9 & 0.020 \\ 
		J1610$+$1811 & 6.0 & 5.11 & 6.8	& 2.3	& 3.3 & 0.013 \\ 
		\hline
	\end{tabular}
    \end{tightcenter}
    {\footnotesize (1) Object name. (2) Projected jet length, in units of arcseconds. (3) Source 0.5--7.0\,keV counts from the jet. (4) Observed 0.5--7.0\,keV jet flux, in units of $10^{-15}\rm\,erg\,cm^{-2}\,s^{-1}$. (5) Observed 0.5--7.0\,keV jet surface brightness, in units of $10^{-15}\rm\,erg\,cm^{-2}\,s^{-1}\,arcsec^{-2}$ (6) Rest-frame 2--10\,keV jet luminosity, in units of $10^{44}\rm\,erg\,s^{-1}$. (7) Ratio of X-ray fluxes for the jet to core. All flux-related measurements have an approximate factor of two uncertainty.}
\end{table}

Using the annular sectors from Section~\ref{sect:annulus}, we determined for each source which X-ray sector(s) was (were) coaligned with a radio feature. In total, we defined 11 sectors from the nine sources that were coaligned with radio features. Counts from the 11 sectors were summed, giving us values for the observed, simulated PSF, and background counts. In total, we measured 16 observed counts and expect 14.37 simulated PSF counts plus 0.87 background counts. Based on our detection criterion from Section~\ref{sect:annulus}, this gives a false detection probability of 0.456. As this does not satisfy our detection threshold of $p\leq$\,0.005, extended X-rays are not detected from the stacked image. 

Given the lack of detection from the stacked analysis, we instead determined the upper limit of the average X-ray jet flux for the nine quasars. We began by estimating an upper limit count rate using the method described in \cite{Kashyap2010}. To detect a source at our defined 0.005 probability for a total background of 15.24 counts (simulated PSF plus field background), we need a minimum of 27 observed counts. We therefore must expect 36.08 counts to ensure that we obtain  $\geq 27$ counts to 95\% probability. In total, we estimated a required net of 20.84 counts over the background, or a rate of $1.99\times10^{-4}\rm\,cts\,s^{-1}$.  

Having determined an upper limit count rate, we modeled the jet with a {\tt phabs$\cdot$powerlaw} expression. The Galactic column density was fixed to the average value of the nine sources, giving us $N_{\rm H} = 3.02\times10^{20}\rm\,cm^{-2}$. Consistent with Section~\ref{sect:flux_extend}, we fixed the jet photon index to \mbox{$\Gamma=1.9$}. Using \pimms{}v4.11a with our count rate limit of $1.99\times10^{-4}\rm\,cts\,s^{-1}$, we found an upper limit for the average unabsorbed jet flux of  $2.26\times10^{-15}\rm\,erg\,cm^{-2}\,s^{-1}$ over the observed 0.5--7.0\,keV band. This estimated flux is consistent with typical background limits for \chandra\ ACIS observations.

With the X-ray jet flux limit now available, we tested for evidence of IC/CMB emission by calculating the average X-ray jet-to-core ratio limit for the nine quasars in our sample with no detected X-ray jets. We measured the jet-to-core flux ratio limit for each quasar and then averaged the values, which we found to be 2.4\%. These results are consistent with the mean 2.2\% flux ratio of the five quasars with X-ray features. The flux ratio is also in agreement with the distribution of jet-to-core ratios found by \cite{Marshall2018}. Overall, these results reinforce the finding of Section~\ref{sect:iccmb} that there is no observed redshift dependence in the X-ray jet emission. 

\subsection{Non-Coincident Radio and X-ray Features in J1016$+$2037}
\label{sect:J1016}

Examination of our quasar sample shows that the majority of X-ray features, when detected, are spatially coincident with radio features. However, J1016$+$2037 stands out amongst our sample due to its non-coincident X-ray and radio features, where the measured offset between the X-ray emission and radio feature centroid is $195^{\circ}$. This offset is despite the excellent alignment of the X-rays and radio from its quasar core. Furthermore, the detection probability for the extended X-rays from J1016 is amongst the highest rated for our sample (Table~\ref{table:extend}), suggesting that this feature is indeed real. Thus, the origin of the X-ray feature and its misalignment with the radio merits further discussion.

We began by verifying that the known \chandra\ PSF artifact (see Section~\ref{sect:observation}) was not aligned with the observed X-ray feature in J1016, confirming that the extended X-rays are not due to a systematic effect from \chandra. We then investigated the possibility of an unassociated X-ray source being the origin of the extended X-rays observed in J1016. Our measured X-ray flux for the extended source is $\sim$\,$10^{-14}\rm\,erg\,cm^{-2}\,s^{-1}$ (Table~\ref{table:flux}), and there are $\sim$\,100 X-ray sources\,deg$^{-2}$ above such a flux limit \citep{Civano2016}. As a result, there is only a $\sim$\,1\% probability of a chance association within 5\arcsec\ of any quasar in our sample. We also confirmed that the X-ray feature is not coincident with any optical source in the Pan-STARRS 3Pi survey to a minimum brightness limit of 21.3\,mag in the $grizy$ bands \citep{Tonry2012}. It is therefore likely that the extended X-rays are associated with J1016. 

Given that the offset between the radio and X-ray features is $\sim$\,$180^{\circ}$, it is possible that the different emissions may originate from jet/counterjet features. One-sided radio emission from quasars is normally interpreted as jet emission, meaning that the X-rays would be from the counterjet side. However, this scenario is unlikely as the beamed jet will generally have a higher X-ray flux than the counterjet region, and so X-rays should be detected coincident with the observed radio feature. In addition, the measured jet-to-core X-ray flux ratio for J1016 (Table~\ref{table:flux}) is consistent with flux ratios from a beamed jet, suggesting that the observed extended X-rays are due to emission from the near jet. It is therefore unclear from the multiwavelength data what is the orientation of the jetted outflow in J1016, assuming jets are present.  

Despite concluding that the extended X-rays in J1016 are associated with the system, we lacked the X-ray count statistics required to determine the physical origin of the observed misalignment. Follow-up, deep-exposure observations with \chandra\  will permit a spectroscopic and morphological analysis of the X-ray structure where different emission models may be investigated to determine the physical origin for this irregular emission feature. 

\section{Optical Properties of Quasar Sample} 
\label{sect:ox}

\begin{table}
	\caption{Optical Properties of the Quasar Sample}
	\label{table:alpha}
	\begin{tightcenter}
	\begin{tabular}{ c c D c }
		\hline \hline
		Object & $m_{1450\rm\,\angstrom}$ & \multicolumn2c{$\ell_{2500\rm\,\angstrom}$} & $\alpha_{\rm ox}$ \\
		(1) & (2) & \multicolumn2c{(3)} & (4) \\
		\hline
		\decimals
        J0801$+$4725 & 19.58 & 6.7 & $-1.61\substack{+0.09\\-0.10}$ \\
		J0805$+$6144 & 19.92 & 4.1 & $-1.35\substack{+0.03\\-0.03}$ \\ 
		J0833$+$0959 & 21.09 & 2.3 & $-1.25\substack{+0.06\\-0.06}$ \\
		J0909$+$0354 & 19.94 & 4.9 & $-1.28\substack{+0.03\\-0.03}$ \\
		J0933$+$2845 & 17.92 & 34.5 & $-1.86\substack{+0.08\\-0.08}$ \\
	    J1016$+$2037 & 19.12 & 9.1 & $-1.51\substack{+0.05\\-0.05}$ \\
	    J1128$+$2326 & 18.56 & 14.5 & $-1.76\substack{+0.07\\-0.07}$ \\
	    J1223$+$5038 & 17.47 & 54.7 & $-1.63\substack{+0.03\\-0.03}$ \\
	    J1405$+$0415 & 19.97 & 4.4 & $-1.37\substack{+0.04\\-0.04}$ \\
	    J1435$+$5435 & 20.19 & 5.5 & $-1.72\substack{+0.13\\-0.14}$ \\
	    J1610$+$1811 & 18.33 & 18.9 & $-1.52\substack{+0.03\\-0.03}$ \\
	    J1616$+$0459 & 19.19 & 9.1 & $-1.44\substack{+0.04\\-0.04}$ \\
	    J1655$+$3242 & 19.58 & 6.3 & $-1.74\substack{+0.11\\-0.12}$ \\
	    J1655$+$1948 & 20.02 & 4.4 & $-1.56\substack{+0.09\\-0.10}$ \\
		\hline
	\end{tabular}
	\end{tightcenter}
	(1) Object name. (2) Rest-frame, monochromatic AB apparent  magnitude at 1450\,\angstrom, as measured from Pan-STARRS $r$ band. (3) Rest-frame, monochromatic luminosity at 2500\,\angstrom, in units of $10^{31}\rm\,erg\,s^{-1}\,Hz^{-1}$. (4) Optical-to-X-ray power-law slope. Fluxes are assumed to have a 0.1\,mag uncertainty, and $\alphaox$ errors include both optical and X-ray uncertainties.
\end{table}

Previous quasar studies have demonstrated an inverse relationship between the optical-to-X-ray flux ratio and optical luminosity \citep[e.g.,][]{Avni1982,Tananbaum1986,Wilkes1994}. We therefore compiled optical properties of our sample in an effort to investigate the optical and X-ray relationship amongst our sources. 

In keeping with our previous quasar analysis \citep{Snios2020b}, we obtained optical fluxes at the rest-frame wavelength of 1450\,\angstrom. For our current sample, this corresponds to observed wavelengths between 5850--7000\,\angstrom. We consequently utilized the monochromatic AB apparent magnitude from the Pan-STARRS $r$ band as it has an effective wavelength of 6241\,\angstrom{} \citep{Tonry2012}. All sources were identified in the Pan-STARRS 3Pi Survey catalog, and no Galactic extinction corrections were applied to our measurements. We additionally extrapolated the rest-frame $m_{1450\rm\,\angstrom}$ results to determine the rest-frame 2500\,\angstrom{} luminosity $\ell_{2500\rm\,\angstrom}$. A UV spectral index of $\alpha = -0.5$ was assumed for our extrapolation ($f_{\nu} \propto \nu^{\alpha}$), which is consistent with prior quasar studies \citep[i.e.][]{Shemmer2006,Nanni2017,Snios2020b}. The optical fluxes and luminosities are shown in Table~\ref{table:alpha}. 

Having determined both the optical and X-ray luminosities for the sample, we calculated the optical-to-X-ray power-law index $\alphaox$ for each source. We defined $\alphaox$ the same as \cite{Tananbaum1979}: 
\begin{equation}
\frac{{\rm log}(\ell_{\rm 2\,keV}/\ell_{\rm 2500\,\angstrom})}{{\rm log}(\nu_{\rm 2\,keV}/\nu_{\rm 2500\,\angstrom})} = 0.3838\cdot\rm log(\ell_{2\,keV}/\ell_{\rm 2500\,\angstrom}),
\end{equation}
where $\ell_{\rm 2\,keV}$ and $\ell_{\rm 2500\angstrom}$ are the monochromatic luminosities at 2\,keV and 2500\,\angstrom, respectively. Consistent with \cite{Snios2020b}, the $\alphaox$ error was derived by summing the X-ray luminosity uncertainty in quadrature with an approximated 10\% optical uncertainty, or $\sim$0.1 mag. Results for $\alphaox$ are provided in Table~\ref{table:alpha}.

\subsection{Optical and X-Ray Relationship}

\begin{figure}
    \begin{tightcenter}
    \includegraphics[width=0.99\linewidth]{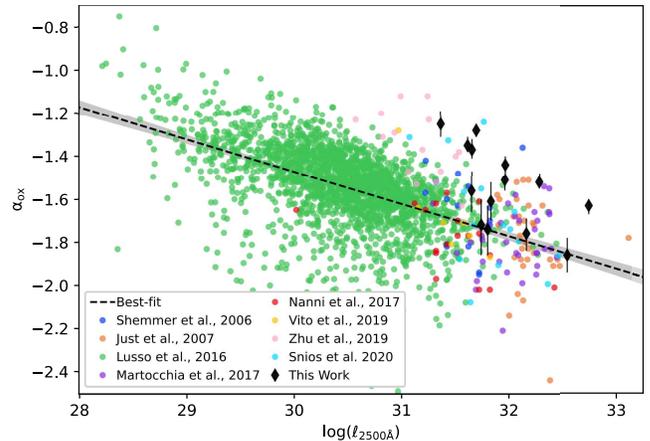}
    \end{tightcenter}
\caption{Optical-to-X-ray power-law slope $\alphaox$ vs. UV luminosity $\ell_{2500\rm\,\angstrom}$. Results from this work (black) are compared against other quasar samples from the literature. The black dotted line is the best-fit model, while the gray region is the $3\sigma$ confidence level. Previous measurements are taken from \cite{Shemmer2006,Just2007,Lusso2016,Siemiginowska2016,Nanni2017,Martocchia2017,Zhu2019,Vito2019b, Snios2020b}.}
\label{fig:alpha}
\end{figure}

\begin{table}
	\caption{Radio Properties of the Sample}
	\label{table:radio}
	\begin{tightcenter}
	\begin{tabular}{ c D D c }
		\hline \hline
		Object & \multicolumn2c{$f_{\rm5\,GHz}$} & \multicolumn2c{$\ell_{\rm5\,GHz}$} & log($\ell_{\rm5\,GHz}$/$\ell_{\rm2\,keV}$)\\
		(1) & \multicolumn2c{(2)} & \multicolumn2c{(3)} & (4) \\
		\hline
        J0801$+$4725 & 89~ & 6.7 & $6.2\pm0.3$ \\
        J0805$+$6144 & $798^{*}$ & 50.4 & $6.6\pm0.2$ \\
        J0833$+$0959 & 150~ & 15.5 & $6.1\pm0.2$ \\
        J0909$+$0354 & 125$^{*}$ & 9.6 & $5.6\pm0.2$ \\
        J0933$+$2845 & $76^{*}$ & 6.4 & $6.1\pm0.3$ \\
        J1016$+$2037 & 808~ & 54.4 & $6.7\pm0.2$ \\
        J1128$+$2326 & 174~ & 11.1 & $6.5\pm0.2$ \\
		J1223$+$5038 & 265~ & 23.7 & $5.9\pm0.2$ \\
        J1405$+$0415 & 1187~ & 85.9 & $6.8\pm0.2$ \\
        J1435$+$5435 & 115~ & 12.5 & $6.8\pm0.5$ \\
        J1610$+$1811 & 239~ & 16.2 & $5.9\pm0.2$ \\
        J1616$+$0459 & $393^{*}$ & 28.6 & $6.2\pm0.2$ \\
        J1655$+$3242 & 201~ & 14.4 & $6.9\pm0.4$ \\
        J1655$+$1948 & $194^{*}$ & 14.6 & $6.6\pm0.3$ \\
		\hline
	\end{tabular}
	\end{tightcenter}
	(1) Object name. (2) Rest-frame 5\,GHz flux density from the FIRST VLA survey \citep{White1997}, in units of $\rm mJy$. In cases where no FIRST detection was found (denoted with a ${}^{*}$), flux densities were measured from 1.4\,GHz VLA observations. (3) Monochromatic luminosity at 5\,GHz, in units of $10^{33}\rm\,erg\,s^{-1}\,Hz^{-1}$.  (4) Rest-frame radio-to-X-ray ratio. Flux densities and luminosities are assumed to have a 15\% uncertainty. Radio-to-X-ray ratio errors include both radio and X-ray uncertainties. 
\end{table}

Prior studies of quasars have shown an anticorrelation between $\alphaox$ and $\ell_{2500\angstrom}$ that is independent of redshift \citep{Bechtold1994,Vignali2003, Steffen2006, Kelly2007, Nanni2017,Snios2020b}, so we investigated if our radio-loud quasar sample was consistent with this relationship. In keeping with the method from \cite{Snios2020b}, we selected quasars from the literature with known optical and X-ray properties.  We included 16 targets from \cite{Shemmer2006}, 34 from \cite{Just2007}, 2153 from \cite{Lusso2016}, 18 from \cite{Nanni2017}, 35 from \cite{Martocchia2017}, 15 from \cite{Zhu2019}, 7 from \cite{Vito2019b}, and 15 from \cite{Snios2020b}. In total, we compiled a sample of 2307 quasars with known $\alphaox$ and $\ell_{2500\angstrom}$ parameters for our analysis. 

The resulting $\alphaox$\,-\,$\ell_{2500\angstrom}$ relationship from the complete dataset is shown in Figure~\ref{fig:alpha}. Using a linear regression with the {\tt scipy} Python package \citep{Virtanen2020}, we found a best-fit relation of 
\begin{equation}
    \label{eq:alpha1}
    \alphaox = (-0.150\pm0.005)\,{\rm log}(\ell_{2500\angstrom}) + (3.0\pm0.1),
\end{equation}
for the total sample, where the errors are reported to 1$\sigma$. Our best fit is consistent with previous studies to within 1$\sigma$ \citep{Nanni2017,Snios2020b}. Additionally, we verified that the best-fit relationship is independent of redshift, which is consistent with previous works \citep[i.e.,][]{Just2007, Snios2020b}. We repeated the analysis using only the 14 quasars from this work, finding a slope of $-0.3\pm0.1$ and an intercept of $9\pm4$, where the best fit was determined using an orthogonal distance regression in {\tt scipy}. Thus, the quasar sample discussed in this paper is consistent within $2\sigma$ to the overall trend for the broader quasar population.

\subsection{Impact of Radio Loudness on Optical Properties}
\label{sect:op-rl}

\begin{figure}
    \begin{tightcenter}
    \includegraphics[width=0.99\linewidth]{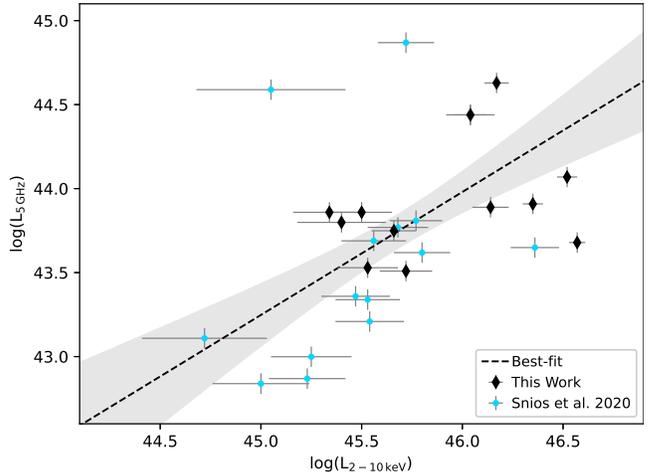}
    \end{tightcenter}
\caption{Comparison of rest-frame radio ($L_{\rm5\,GHz}$) and X-ray ($L_{\rm 2-10\,keV}$) luminosities. The black dotted line is the best-fit relation, while the gray region is the $1\sigma$ confidence level. Radio-to-X-ray ratios are broadly consistent for our sample.}
\label{fig:radio}
\end{figure}

Radio-loud quasars generally have a higher X-ray luminosity than radio-quiet quasars for a given optical luminosity \citep[e.g.,][]{Worrall1987, Wu2013, Zhu2019}. Given that our $\alphaox$\,-\,$\ell_{2500\angstrom}$ analysis includes both radio-loud and radio-quiet quasars, we investigated the impact of radio loudness on the $\alphaox$\,-\,$\ell_{2500\angstrom}$ relationship. Our sample of 2307 quasars was separated into radio-loud and radio-quiet populations, where the radio-loud dataset includes all sources from \cite{Zhu2019}, \cite{Snios2020b}, and this work. This gave us a total 44 radio-loud sources and 2263 radio-quiet sources for our analysis. 

Using the radio-loud dataset, we found a linear relationship between $\alphaox$\,-\,$\ell_{2500\angstrom}$ with a best-fit slope of $-0.33\pm0.06$ and an intercept of $9.1\pm1.8$. Repeating our analysis with the radio-quiet sample, we found a slope of $-0.157\pm0.005$ and an intercept of $3.2\pm0.1$. The radio-loud sample therefore diverges from the radio-quiet sample by $\sim$\,$3\sigma$. 
Because we previously verified that our best fit was independent of redshift, the observed discrepancy between the radio-loud and radio-quiet sample is not due to differences in redshift selection criteria.

Despite the observed impact of radio-loudness on the $\alphaox$\,-\,$\ell_{2500\angstrom}$ relationship, we stress that this result may be biased and/or incomplete due to our small sample size of radio-loud quasars. The available radio-loud quasar sample lacks the comprehensive coverage of X-ray and UV luminosities needed to accurately fit a relationship for a broad range of sources. Further study with a larger radio-loud sample is required in order to verify the dependence of radio loudness on $\alphaox$ properties. 

\section{Radio Properties of Quasar Sample} 
\label{sect:radio}

Given that our quasar sample is comprises of sources detected in both X-rays and radio, we investigated the relation between their radio and X-ray luminosities. Radio flux densities for quasars discussed in this work were provided from the VLA-FIRST survey \citep{White1997}, which observed at 1.4\,GHz. In cases where no FIRST detection was found, flux densities were directly measured from archival 1.4\,GHz VLA observations. For our sample, these radio measurements correspond to rest-frame frequencies between 5.6--6.7\,GHz, which we assumed to be a reasonable approximation of the rest-frame 5\,GHz flux density. Radio flux densities and luminosities for the sample are provided in Table~\ref{table:radio}. We note that the reported radio luminosities represent the core emission for each source. 

The quasars from this sample were added together with the radio-loud sources from \cite{Snios2020b}, giving us 29 radio-loud quasars at $z>3$ with known radio and X-ray fluxes. The radio--X-ray luminosity relationship for these sources is shown in Figure~\ref{fig:radio}. We fit the data using linear orthogonal data regression in {\tt scipy}, and we found a best-fit relation of
\begin{equation}
{\rm log}(L_{\rm5\,GHz}) = (0.73\pm0.22){\rm log}(L_{\rm2-10\,keV}) + (10.3\pm10.3),
\end{equation}
where the reported errors are $1\sigma$. We note that outliers from \cite{Snios2020b} that reside above the best fit are known to be either Compton-thick and/or possess unresolved X-ray structure and are likely not representative of the remaining sample. In comparison, all sources from our current work agree with the measured relationship within 3$\sigma$. 
Our measured radio--X-ray luminosity correlation is consistent with the relationship for radio-loud quasars at $z < 2$ \citep{Fan2016}, further reinforcing the lack of spectral evolution over redshift. Additional, in-depth analysis of the radio and X-ray properties of these quasars and fundamental plane results, which is beyond the scope of this work, will be discussed in a forthcoming paper.

\section{Conclusions}
\label{sect:conclude}
 
We analyzed \chandra\ observations of 14 radio-loud quasars at redshifts $3 < z < 4$, each with radio features in addition to core emission, to measure their X-ray spectral properties and search for evidence of resolved structure. We detected all quasars in the 0.5--7.0\,keV band and extracted emission spectra of the quasar cores. Each spectrum was fit with an absorbed power-law model, where our mean best-fit photon index is \mbox{$1.6\pm0.2$}. Observed X-ray fluxes and rest-frame luminosities were also determined from the spectral best fits. Additionally, we detected an Fe\,K emission line from the quasar J1223$+$5038 at high significance. 

We performed a morphological analysis of each X-ray source using \chandra\ observations, and we detected X-ray features at distances up to 12\arcsec{} from the quasar core in five of the sources. The X-ray features are spatially coincident with existing radio features for four of the five sources, suggesting that the majority of the X-ray features are jets. J1016$+$2037 stands out amongst our sample due to a $\sim$\,$180^{\circ}$ misalignment between its X-ray and radio features. We speculated on the cause this observed misalignment, but the available X-ray counts were insufficient to conclusively determine its origin. 

Rest-frame 2--10\,keV luminosities of the X-ray jets were estimated for the five quasars with extended X-rays, and their X-ray jet-to-core flux ratios were measured to be up to 3.6\%. We also estimated an upper limit on the average X-ray jet flux for the remaining nine quasars from a stacked image analysis, finding a limit of $2.3\times10^{-15}\rm\,erg\,cm^{-2}\,s^{-1}$. This flux limit corresponds to an average jet-to-core X-ray flux ratio upper limit of 2.4\%. Our measured jet-to-core flux ratios, both for those directly measured and the upper limits, agree well with measurements from low-redshift quasars, suggesting that the observed X-ray jet emission mechanism is independent of redshift. Deeper \chandra\ observations are required to investigate the spectrum and spatial structure of the detected X-ray features. 

Beyond our morphological study, we determined the optical-to-X-ray power-law slope $\alphaox$ for each quasar core using optical/UV data available in the literature. We observed a clear anticorrelation trend between $\alphaox$ and $\ell_{2500\angstrom}$, where our derived best-fit relationship is consistent with other quasar surveys. We also measured radio-to-X-ray luminosity ratios for our sources, and our results are broadly consistent with other radio-loud quasar surveys regardless of redshift. These multiwavelength results reinforce that the spectral evolution of quasars is independent of redshift.  

Our results demonstrate the strength of high-resolution X-ray imaging in studying quasars at high redshifts. Further sampling of the high-redshift quasar population with high-resolution instruments, such as \chandra\ or the proposed missions AXIS \citep{Mushotzky2019} and Lynx \citep{Gaskin2019}, will allow us to investigate jet properties over a broad redshift range while also identifying extended X-ray features ideal for follow-up morphological and spectroscopic analyses.

\acknowledgements{
We thank the referee for their comments and suggestions. B.S., D.A.S., A.S., and M.S. were supported by NASA contract NAS8-03060 (\chandra\ X-ray Center). B.S. and D.A.S. were also supported by CXC grant GO8-19077X. Work by C.C.C. at the Naval Research Laboratory is supported by NASA DPR S-15633-Y. Radio observations were provided by the National Radio Astronomy Observatory, a facility of the National Science Foundation operated under cooperative agreement by Associated Universities, Inc.

\software{\ciao v4.12, \caldb v4.9.2.1 \citep{Fruscione2006}, Sherpa \citep{Freeman2001}, {\tt scipy} \citep{Virtanen2020}}
}

\bibliographystyle{aasjournal}
\bibliography{all_data}

\end{document}